\documentclass[aps,prd,noshowpacs,preprintnumbers,nofootinbib,floatfix,twoside]{revtex4}

\usepackage{graphicx}  % standard LaTeX graphics tool;
\usepackage{amsmath}   % For getting proper math eqns
\usepackage{amssymb}   % Used for getting various symbols eg. gtrsim, lesssim etc.

\def\eq#1{{Eq.~(\ref{#1})}}

\newcommand{\cc}{cosmological constant}

\newcommand{\Cal}[1]{\ensuremath{\mathcal{#1}}}

\newcommand{\LL}{Lanczos-Lovelock }
\newcommand{\D}{\ensuremath{\nabla}}

\newcommand{\LDm}{\ensuremath{\Cal{L}^{(D)}_m}}
\newcommand{\eqn}[1]{Eq.\eqref{#1}}
\newcommand{\ph}[1]{\phantom{#1}}

\def\frab#1#2{\left(\frac{#1}{#2}\right)} 

\begin{document}

\title{DARK ENERGY AND ITS IMPLICATIONS FOR GRAVITY}

\author{T. Padmanabhan}
\email{paddy@iucaa.ernet.in}
\affiliation{IUCAA, Post Bag 4, Ganeshkhind, Pune - 411 007, India\\}

\date{\today}

%%%%%%%%%%%%%%%%%%%%%%%%%%%%%%%%%%%%%ABSTRACT%%%%%%%%%%%%%%%%%%%%%%%%%%%%%%%%%%%%%%%%
\begin{abstract}
The \cc\ is the most economical candidate for dark energy.  No other approach really alleviates the difficulties faced by the \cc\ because, in all other attempts to model the dark energy, one still has to explain why the bulk \cc\ (treated as a low-energy parameter in the action principle) is zero.  I argue that 
until the theory is made invariant under the shifting of the Lagrangian by a constant, one cannot obtain a satisfactory solution
to the cosmological constant problem. This is impossible in any generally covariant theory with the conventional low-energy matter action, if the metric is varied in the action to obtain the field equations. I  review an alternative  perspective  in which gravity arises as  an emergent,
long wavelength phenomenon,   and can be described in terms of an effective theory using an action associated with  null vectors in the spacetime. This action is explicitly invariant under the shift of the  energy momentum tensor $T_{ab}\to T_{ab}+\Lambda g_{ab}$ and any bulk \cc\ can be gauged away. Such an approach seems to be necessary for addressing the \cc\ problem and can easily explain why its bulk value is zero. I describe some possibilities for obtaining its observed value from quantum gravitational fluctuations.

\end{abstract}

\maketitle

\section{Why do we believe in dark energy?}

The simplest possible universe one could imagine will contain just baryons and radiation. However, host of astronomical observations available since mid-70s indicated that the bulk of the matter in the universe is nonbaryonic and dark. Around the same time, the theoretical prejudice for $\Omega_{tot}=1$ gained momentum,\footnote{It is convenient to measure the 
energy densities of the different species, which drive the expansion of the universe, in terms of this \textit{critical density} using the dimensionless parameters $\Omega_i=\rho_i/\rho_c$ (with $i$ denoting
the different components like baryons, dark matter, radiation, etc.) The critical energy density is defined as  $\rho_c=3H^2_0/8\pi G$ where $H_0=\dot a/a$ is the expansion rate of the universe.}  largely led by the inflationary paradigm \cite{inflation}. During the eighties, this led many theoreticians to push (as usual wrongly!) for a model  with $\Omega_{tot}\approx\Omega_{DM}\approx1$ in spite of the fact that host of astronomical observations demanded that $\Omega_{DM}\simeq 0.2-0.3$. 
The indications that the universe indeed has another component of energy density --- so that $\Omega_{tot}=1$ can be reconciled with $\Omega_{DM}\simeq 0.2-0.3$ --- started accumulating in the late eighties and early nineties. Early analysis of several observations \cite{earlyde} indicated that this component is unclustered and has negative pressure. This is confirmed dramatically by the supernova observations in the late nineties (see Ref.~\cite{sn}; for a critical look at the  data, see e.g., Ref.~\cite{tptirthsn1}).  The observations suggest that the missing component contributes  $\Omega_{DE}\cong 0.60-0.75$ and has an \textit{equation-of-state parameter}
$w\equiv p/\rho\lesssim-0.78$. Treated as a fluid, this component has negative pressure (assuming positive energy density) and has been dubbed as \textit{dark energy}. 

 The simplest choice for such dark energy with negative pressure is the cosmological constant which is  a term that can be added to Einstein's equations. This term \textit{acts like  a fluid} with an equation of state $p_{DE}=-\rho_{DE}$.
Combining this with all other observations \cite{cmbr,baryon,h,bao}, we end up with a rather strange composition for the universe with
 $0.98\lesssim\Omega_{tot}\lesssim 1.08$  in which radiation (R), baryons (B), dark matter, made of weakly interacting massive particles (DM) and dark energy (DE) contributes  $\Omega_R\simeq 5\times 10^{-5},\Omega_B\simeq 0.04,\Omega_{DM}\simeq 0.26,\Omega_{DE}\simeq 0.7,$ respectively. 
 
 The key observational feature of dark energy --- which dominates over everything else today in such a universe ---  is that it leads to an accelerated expansion of the universe. When treated as a fluid with a stress tensor $T^a_b=$ dia     $(\rho, -p, -p,-p)$, 
it has an equation state $p=w\rho$ with $w \lesssim -0.8$ at the present epoch. 
 In general relativity,  the source of geodesic  acceleration is $(\rho + 3p)$ and not $\rho$.
  As long as $(\rho + 3p) > 0$, gravity remains attractive while $(\rho + 3p) <0$ can
  lead to `repulsive' gravitational effects. In other words, dark energy with sufficiently negative pressure will
  accelerate the expansion of the universe, once it starts dominating over the normal matter.  This is precisely what is established from the study of high redshift supernova, which can be used to determine the expansion
rate of the universe in the past \cite{sn,snls}.

While most physicists will look at such a weird composition for the universe  with the suspicion it deserves, cosmologists have unhesitatingly accepted such a `concordance model' over the last one decade. The reason is very simple. The concordance  model  is mandated by a host of  observations  and the cosmological paradigm based on such a composition is  remarkably successful. This paradigm is a complex mix of several ingredients and works \cite{adcos} broadly as follows:

(a) The basic  idea is that if  small fluctuations in the energy density existed in the early universe, then gravitational instability can amplify them   leading to structures like galaxies etc. which exist today. The popular procedure for generating these fluctuations is based on the idea that if the very early universe went through an inflationary phase \cite{inflation}, then the quantum fluctuations of the field driving the inflation can lead to energy density fluctuations \cite{genofpert,tplp}. 

(b) While the inflationary models are far from unique and hence lacks predictive power, it is certainly possible to construct models of inflation such that these fluctuations are described by a Gaussian random field and are characterized by a power spectrum of the form $P(k)=A k^n$ with $n\simeq 1$. The inflationary models cannot predict the value of the amplitude $A$ in an unambiguous manner. But it can be determined from CMBR observations and the inflationary model parameters can be fine-tuned to reproduce the observed value. The CMBR observations are consistent with the inflationary model for the generation of perturbations and gives $A\simeq (28.3 h^{-1} Mpc)^4$ and $n\lesssim 1$. (The first results were from COBE \cite{cobeanaly} and
WMAP etc. \cite{cmbr} has re-confirmed them with far greater accuracy). 

(c) One can evolve the initial perturbations by a well understood linear perturbation theory
when the perturbation is small. But when $\delta\approx(\delta\rho/\rho)$ is comparable to unity the perturbation theory
breaks down and one has to resort to numerical  simulations \cite{baryonsimulations}
 or theoretical models based on approximate 
 ansatz \cite{nlapprox, nsr} to understand their evolution --- especially the baryonic part, that leads to observed structures in the universe. 
 
 The resulting model,  which is characterized essentially by seven numbers [ $h\approx 0.7$ describing the current rate of expansion; $\Omega_{DE}\simeq 0.7,\Omega_{DM}\simeq 0.26,\Omega_B\simeq 0.04,\Omega_R\simeq 5\times 10^{-5}$ giving the composition of the universe; the amplitude $A\simeq (28.3 h^{-1} Mpc)^4$ and the index $n\simeq 1$ of the initial perturbations]
seems to be quite successful in explaining the observations. 
This is a nontrivial measure of success since  the paradigm could have been falsified by observations on several counts. For example, the values of $\Omega_Bh^2$ could be constrained by CMBR observations as well as from the deuterium abundance from big bang nucleosynthesis. These two probe the universe at widely different epochs (a few hundred thousand years after big bang compared a few minutes) using very different techniques. The results, nevertheless lead to the same number.
Such tests have led to faith in the concordance model including the existence of dark energy. So, even though we do not understand our universe, we have been quite successful in parametrising our ignorance in terms of well-chosen numbers.

 \section{What if dark energy is just \cc\ ?}
 
 After such a brief overview, I will concentrate  on the dark energy and the issues it rises  (for a few of the recent reviews, see ref. \cite{cc}). The concordance model, as  defined above, uses the fact that
 the simplest model for  a fluid with negative pressure is \textit{not a fluid at all} but the
cosmological constant with $w=-1,\rho =-p=$ constant.  The cosmological constant  introduces a fundamental length scale in the theory $L_\Lambda\equiv H_\Lambda^{-1}$, related to the constant dark energy density $\rho_{_{\rm DE}}$ by 
$H_\Lambda^2\equiv (8\pi G\rho_{_{\rm DE}}/3)$.
Though, in classical general relativity,
    based on  $G, c $ and $L_\Lambda$,  it
  is not possible to construct any dimensionless combination from these constants, when one introduces the Planck constant, $\hbar$, it is  possible
  to form the dimensionless combination $\lambda=H^2_\Lambda(G\hbar/c^3) \equiv  (L_P^2/L_\Lambda^2)$.
  Observations then require $(L_P^2/L_\Lambda^2) \lesssim 10^{-123}$ requiring enormous fine tuning.\footnote{This is, of course, the party line. But it might help to get some perspective on how enormous, the `enormous' really is. To begin with note that, the sensible particle physics
  convention considers ratios of length/energy scales and not their \textit{squares}. This leads to
  $(L_P/L_\Lambda) \sim 10^{-61}$. In standard model of particle physics the ratio between Planck scale to neutrino mass scale is $10^{19}GeV/10^{-2}eV\sim 10^{30}$ for which we have no theoretical explanation. So when we worry about the fine tuning of cosmological constant without expressing similar worries about standard model of particle physics, we are essentially assuming that $10^{30}$ is not a matter for concern but $10^{61}$ is. This subjective view is defensible but needs to be clearly understood.}

  In the earlier days, this was considered puzzling but 
most people believed that this number $\lambda$ is actually zero. The \cc\ problem in those days was to understand
why it is strictly zero. Usually, the vanishing of a   
constant (which could have appeared in the low energy sector of the theory) indicates an underlying symmetry of the 
theory. For example, the vanishing of the mass of the photon is closely related to
the gauge invariance of electromagnetism. No such symmetry principle is known to operate
at low energies which made this problem very puzzling. There is a symmetry --- called
supersymmetry --- which does ensure that $\lambda =0$ but it is known that supersymmetry is 
broken at sufficiently high energies and hence cannot explain the observed value of $\lambda$.

Given the observational evidence for  dark energy in the universe and the fact that
the simplest candidate for dark energy, consistent with all observations today, is a cosmological 
constant with $\lambda \approx 10^{-123}$ the cosmological constant problem has got linked to the 
problem of dark energy in the universe. So, if we accept  the simplest  interpretation of the
current observations, we need to explain why cosmological constant is non zero
and has this small value. It should, however, be stressed that these 
 two --- the cosmological constant problem and the explaining of dark energy ---
 are 
logically independent issues. \textit{Even if all the observational evidence for dark energy goes
away we still have a \cc\ problem --- viz., explaining why $\lambda$ is zero.}

There is another, related, aspect to \cc\ problem which need to be stressed. In conventional approach to gravity, one derives the equations of motion
from a Lagrangian $\mathcal{L}_{\rm tot} = \mathcal{L}_{\rm grav}(g) + \mathcal{L}_{\rm matt}(g,\phi)$ where
$\mathcal{L}_{\rm grav}$ is the gravitational Lagrangian dependent on the metric and its derivative
and $\mathcal{L}_{\rm matt}$ is the matter Lagrangian which depends on both the metric and the 
matter fields, symbolically denoted as $\phi$. In such an approach, the cosmological constant can be introduced via two different routes
which are conceptually different but operationally the same. 
First, one may decide
to take the gravitational Lagrangian to be $\mathcal{L}_{\rm grav} =(2\kappa)^{-1}(R-2\Lambda_g)$
where $\Lambda_g$ is a parameter in the  (low energy effective) action  just like
the Newtonian gravitational constant $\kappa$. 
The second route  is by
shifting  the matter Lagrangian by $\mathcal{L}_{\rm matt}\to \mathcal{L}_{\rm matt} - 2\lambda_m$. Such a shift is clearly
equivalent to adding a cosmological constant $2\kappa\lambda_m$ to the
$\mathcal{L}_{\rm grav}$. In general, what can be observed through gravitational interaction 
is the combination $\Lambda_{\rm tot} = \Lambda_g
+ 2\kappa\lambda_m$.  

It is now clear that there are two distinct aspects to the  cosmological
constant problem. The first question is why $\Lambda_{\rm tot} $ is very small
when expressed in natural units. Second, since $\Lambda_{\rm tot}$ could have
had two separate contributions from the gravitational and matter sectors, why
does the \textit{sum} remain so fine tuned? This question is particularly relevant because it is believed that our universe went through several phase transitions in the course of its  evolution, each of which shifts the energy momentum tensor of matter by $T^a_b\to T^a_b+L^{-4}\delta^a_b$ where $L$ is the scale characterizing the transition. For example, the GUT and Weak Interaction scales are about $L_{GUT}\approx 10^{-29}$ cm, $L_{SW}\approx 10^{-16}$ cm respectively which are tiny compared to $L_{\Lambda}$. 
Even if we take a more pragmatic approach, the observation of Casimir effect in the lab sets a bound that $L<\mathcal{O}(1)$ nanometer, leading to a $\rho$ which is about $10^{12}$ times the observed value \cite{gaurang}. 

Finally, I will comment on two other issues related to \cc\ which appear frequently in the literature.
The first one is what could be called the ``why now'' problem of the \cc. 
How come the energy density contributed by  the \cc\ (treated as the dark energy) is comparable to the energy density of the rest of the matter at the \textit{current epoch} of the universe? I  do not believe this is an \textit{independent} problem; 
if we have a viable theory predicting a particular numerical value for $\lambda$, then
the energy density due to this \cc\ will be comparable to the rest of the energy density at
\textit{some} epoch. So the real problem is in understanding the numerical value of $\lambda$;
once that problem is solved the `why now' issue will take care of itself. In fact, we do not have a viable
theory to predict the current energy densities of \textit{any} component  which populates the universe, let alone the dark energy!.
For example, the energy density of radiation today is computed from its temperature which is an 
observed parameter --- there is no theory which tells us that this temperature has to be
2.73 K when, say,  galaxy formation has taken place for certain billion number of years. Neither do we have a theory which predicts the value of $\Omega_R/\Omega_B$ at the present epoch. So the really important issue is to fix the numerical value of $L_\Lambda$ in terms of Planck length.\footnote{For example, if  some nonperturbative quantum gravity effect involving the exponential of a semiclassical action --- or something similar ---  lead to a perfectly reasonable looking factor $L_\Lambda/L_P=\exp(\sqrt{2}\pi^4)\approx 10^{60}$, then all the issues are resolved.}

One also notices in the   literature a discussion of the contribution of 
the zero point energies of the quantum fields to the \cc\ which is often  misleading, if not incorrect. What is usually done is to attribute a zero-point-energy $(1/2)\hbar\omega$ to each mode of the field and add up all these energies with an ultra violet cut-off. For an electromagnetic field, for example, this will lead to an integral proportional to
\begin{equation}
\rho_0=\int_0^{k_{max}} dk\ k^2 \hbar k\propto k_{max}^4
\end{equation}  
which will give $\rho_0\propto L_P^{-4}$ if we invoke a Planck scale cut-off with $k_{max}=L_P^{-1}$. It is then claimed that, this $\rho_0$ will contribute to the cosmological constant. There are several problems with such a naive analysis. First, the $\rho_0$ computed above can be easily eliminated by the normal ordering prescription in quantum field theory and what one really should compute is the \textit{fluctuations} in the vacuum energy --- not the vacuum energy itself. Second, even if we take the nonzero value of $\rho_0$ seriously, it is not clear this has anything to do with a \cc. The energy momentum tensor due to the \cc\ has a very specific form $T^a_b\propto\delta^a_b$ and its trace is nonzero. The electromagnetic field, for example, has a stress tensor with zero trace, $T^a_a=0$; hence in the vacuum state
the expectation value of the trace, $\langle \mathrm{vac}|T^a_a|\mathrm{vac}\rangle$, will vanish, showing that
 the equation of state of the bulk electromagnetic vacuum is still
 $\rho_0=3p_0$ which does not lead to a \cc\ . (The trace anomaly will not work in the case of electromagnetic field.) So the naive calculation of vacuum energy density with a cutoff and the claim that it contributes to \cc\ is not an accurate statement in many cases. 
  
\section{What if dark energy is not the \cc\ ?} 

Based on some of these misgivings about the \cc\ many people have tried to come up with alternative explanations for the dark energy. These attempts can be divided into two broad categories. The first set of ideas
 --- which accounts for the bulk of the published research ---
 assumes that the \cc\ is zero for reasons unknown, and invokes some exotic physics (usually using scalar fields, higher dimensional models etc.).  The second set also assumes that the \cc\ is zero for reasons unknown and tries to explain the cosmological observations by some conventional, less esoteric physics. 

The first set of approaches is conceptually no better compared to \cc\ and it is very doubtful whether this ---  rather popular ---  approach, based on scalar fields, has helped us to understand the nature of the dark energy
  at any deeper level. These
  models, viewed objectively, suffer from several shortcomings:
 The most serious problem with them is that they have no predictive power. As can be explicitly demonstrated, virtually every form of expansion history $a(t)$ can be modeled \cite{tptachyon,ellis} by a suitable ``designer" scalar field potential $V(\phi)$.
 What is more,  the scalar field potentials  used in the literature have no natural field theoretical justification. All of them are non-normalizable in the conventional sense and have to be interpreted as a low energy effective potential in an ad hoc manner.
 Observationally, one key difference between \cc\ and scalar field models is that the latter lead to a $(p/\rho) \equiv w(a)$ which varies with time. So they are worth considering if the  observations have suggested a varying $w$, or if observations have ruled out $w=-1$ at the present epoch. However, all available observations are consistent with \cc\ ($w=-1$) and --- in fact --- the possible variation of $w$ is strongly constrained \cite{jbp}. 
  Further, it can be shown that  even when $w(t)$ is determined by observations, it is not possible to proceed further and determine
  the nature of the scalar field Lagrangian. (See the first paper in ref.\cite{tptirthsn1} for an explicit example of such a construction.)\footnote{As an aside, let us note that in drawing conclusions from the  observational data, one should be careful about the hidden assumptions in the statistical analysis. Claims regarding  $w$ depends crucially on the data sets used, priors which are assumed and possible parameterizations which are adopted. (For more details related to these issues, see the last reference in \cite{jbp}.) It is fair to say that all currently available data is consistent with $w=-1$. Further, there is some amount of tension between WMAP and SN-Gold data with the recent SNLS data \cite{snls} being more concordant with WMAP than the SN Gold data.} 
  
Let us next consider the second set of approaches in which --- again --- we assume that the \cc\ is zero because of some unknown reason but try to explain the observed acceleration of the universe in terms of reasonably conservative physics.  
One of the \textit{least} esoteric ideas in this direction
is that the cosmological constant term in the  equations arises because we have not calculated the energy density driving the expansion of the universe correctly. 

This idea arises as follows:  The energy momentum tensor of the real universe, $T_{ab}(t,{\bf x})$ is inhomogeneous and anisotropic. If  we could solve the exact Einstein's equations
$G_{ab}[g]=\kappa T_{ab}$ with it as the source we will be led to a  complicated metric $g_{ab}$. 
The metric describing the large scale structure of the universe should be obtained by averaging this exact solution over a large enough scale, leading to $\langle g_{ab}\rangle $. But since we cannot solve  exact Einstein's equations, what we actually do is to average the stress tensor {\it first} to get $\langle T_{ab}\rangle $ and {\it then} solve Einstein's equations. But since $G_{ab}[g]$ is  nonlinear function of the metric, $\langle G_{ab}[g]\rangle \neq G_{ab}[\langle g\rangle ]$ and there is a discrepancy. This is most easily seen by writing
\begin{equation}
G_{ab}[\langle g\rangle ]=\kappa [\langle T_{ab}\rangle  + \kappa^{-1}(G_{ab}[\langle g\rangle ]-\langle G_{ab}[g]\rangle )]\equiv \kappa [\langle T_{ab}\rangle  + T_{ab}^{corr}]
\end{equation}
If --- based on observations --- we take the $\langle g_{ab}\rangle $ to be the standard Friedman metric, this equation shows that it has, as its  source,  \textit{two} terms:
The first is the standard average stress tensor and the second is a purely geometrical correction term
$T_{ab}^{corr}=\kappa^{-1}(G_{ab}[\langle g\rangle ]-\langle G_{ab}[g]\rangle )$ which arises because of nonlinearities in the Einstein's theory that  leads to $\langle G_{ab}[g]\rangle \neq G_{ab}[\langle g\rangle ]$. If this term can mimic the \cc\ at large scales there will be no need for dark energy and --- as a bonus --- one will solve the ``why now'' problem!
 
 To make this idea concrete, we have to identify an  effective expansion factor $a_{eff}(t)$
 of an inhomogeneous universe (after suitable averaging), and determine the equation of motion satisfied by it. The hope is that it will be sourced by terms so as to have $\ddot a_{eff}(t)>0$ while the standard matter (with $(\rho + 3p)>0$) leads to deceleration of standard expansion factor $a(t)$. Since any correct averaging of
 positive quantities  in $(\rho + 3p)$ will not lead to a negative quantity, the real hope is in defining $a_{eff}(t)$ and obtaining its dynamical equation such that
 $\ddot a_{eff}(t)>0$.  In spite of some recent attention this idea has received \cite{flucde} it is doubtful whether it will lead to the correct result when implemented properly. The reasons for my skepticism are the following:
 
 It is, of course,  obvious that $T_{ab}^{corr}$ is --- mathematically speaking --- non-zero (for an explicit computation, in a completely different context of electromagnetic plane wave, see \cite{gofemw}); the real question  is how big is it compared to $T_{ab}$. When properly done, it seems unlikely that  we will get a large effect for the simple reason that the amount of mass which is contained in the nonlinear regimes in the universe today is subdominant. 
 Any calculation in linear theory or any calculation in which special symmetries are invoked will be inconclusive and untrustworthy in settling this issue. (Several papers on LTB models with mutually contradictory results can be cited as evidence for this!)
 There is also a serious issue  of identifying a suitable analogue of expansion factor from an averaged geometry, which is nontrivial and it is not clear that the answer will be unique. To illustrate this point by an extreme  example,
 suppose we decide to call $a(t)^n$ with, say $n>2$ as the effective expansion factor i.e.,  $a_{\rm eff}(t)=a(t)^n$; obviously $\ddot a_{\rm eff}$ can be positive (`accelerating universe') even with 
 $\ddot a$ being negative. So, unless one has a \textit{unique} procedure to identify the expansion factor of the average universe, it is difficult to settle the issue.
Finally this approach is   strongly linked to explaining the acceleration as observed by SN. Even if we decide to completely ignore all SN data, we still have reasonable evidence for dark energy and it is not clear how this approach can tackle such evidence.

Another equally conservative explanation for the cosmic acceleration will be that we are located in a large underdense region in the universe; so that, locally, the underdensity acts like negative mass and produces a repulsive force. While there has been some discussion in the literature \cite{Hbubble} as to whether observations indicate such a local `Hubble bubble', this does not seem to be a tenable explanation that one can take seriously at this stage. 
For one thing, it is not clear whether such a model --- in which acceleration arises from a local underdensity --- can be reconciled with baryonic acoustic oscillations \cite{bao} and the measured value of Hubble constant \cite{h}. (But whether we are embedded in a local void or not is an important question we should find the answer to --- purely observationally --- irrespective of whether it explains \textit{all} of the observed cosmic acceleration.)

Finally, note that, for any of these ideas to work (scalar field models or more conventional ones),
we first need to find a mechanism which will make the \cc\ vanish. All the scalar field potentials require fine tuning of the parameters in order to be viable. The same comment also applies to the more conventional approaches discussed above. Given this situation, it is certainly worthwhile to consider alternative paradigms in which one has a hope for explaining why \cc\ is zero. Then, one can hope to get the small value of the \cc\ from possibly quantum gravitational considerations.
This is what we will discuss next.

\section{The \cc\ problem demands an alternative perspective on gravity}

Even if all the evidence for dark energy disappears within a decade, \textit{we still need to understand why \cc\ is zero }and much of what I have to say in the sequel will remain relevant. I stress this because there is a recent tendency to forget the fact that the problem of the \cc\ existed (and was recognized as a problem) long before the observational evidence for dark energy, accelerating universe etc cropped up. In this sense, \cc\ problem has an important theoretical dimension which is distinct from what has been introduced by the observational evidence for dark energy.  So,  
it is worth examining this idea in detail and ask how its `problems' can be tackled.
 I will now  argue that the \cc\ problem arises essentially because of our misunderstanding of the nature of gravity and that its solution \textit{demands} an alternative perspective in which the metric tensor is not a dynamical variable and  gravity is treated as an emergent phenomenon -- like elasticity \cite{elastic}. In the later sections I will explicitly describe such a model.

In the  conventional approach to gravity, the gravitational field equations are obtained 
from a Lagrangian $\mathcal{L}_{\rm tot} = \mathcal{L}_{\rm grav}(g) + \mathcal{L}_{\rm matt}(g,\phi)$ by varying the metric, where
$\mathcal{L}_{\rm grav}$ is the gravitational Lagrangian dependent on the metric and its derivative
and $\mathcal{L}_{\rm matt}$ is the matter Lagrangian which depends on both the metric and the 
matter fields, symbolically denoted as $\phi$. This total Lagrangian is integrated
over the spacetime volume with the covariant measure $\sqrt{-g} d^4x$ to obtain the 
action. 

Suppose we now add a constant $(- 2\lambda_m)$ to the matter Lagrangian thereby inducing the change
 $\mathcal{L}_{\rm matt}\to \mathcal{L}_{\rm matt} - 2\lambda_m$. The equations
of motion for matter are invariant under such a transformation which implies that --- in the 
absence of gravity --- we cannot determine the value of $\lambda_m$. 
The transformation $\mathcal{L}\to \mathcal{L}_{\rm matt} - 2\lambda_m$  is a symmetry
of the matter sector (at least at scales below the scale of supersymmetry breaking; we shall ignore supersymmetry in what follows). But,
in the conventional approach, gravity breaks this
symmetry.  \textit{This is the root cause of the  cosmological constant problem.}
As long as
gravitational field equations are of the form $E_{ab} = \kappa T_{ab}$ where $E_{ab}$ is some geometrical quantity (which is $G_{ab}$ in Einstein's theory) the theory
cannot be invariant under the shifts of the form $T^a_b \to T^a_b +\rho \delta^a_b$.
Since such shifts are allowed by the matter sector, it is very difficult to imagine a definitive  solution
to cosmological constant problem within the conventional approach to gravity.

More precisely, consider any model of gravity satisfying the following three conditions: (1)
The metric  is varied in the action to obtain the equations of motion. (2) We demand full general covariance of the equations of motion. (3) The equations of motion for matter sector is invariant under the addition of a constant to the matter Lagrangian.  Then, we can prove `no-go' theorem that the \cc\ problem cannot be solved in such model \cite{gr06}. The proof is elementary. Our demand (2) of general covariance requires the matter action to be an integral over $\mathcal{L}_{matter}\sqrt{-g}$. 
The demand (3) now allows us to add a constant $\Lambda$, say, to $\mathcal{L}_{matter}$ leading to a coupling $\Lambda\sqrt{-g}$ between $\Lambda$ and the metric $g_{ab}$. By our demand (1), when we vary $g_{ab}$ the theory will couple to $\Lambda$ through a term proportional to $\Lambda g_{ab}$ thereby introducing an arbitrary \cc\ into the theory.

The power of the above `no-go theorem' lies in its simplicity! It clearly shows that we cannot solve \cc\ problem unless we drop one of the three demands listed in the above paragraph. Of these, we do not want to sacrifice general covariance encoded in (2); neither do we have a handle on low energy matter Lagrangian so we cannot avoid (3). So the only hope we have is to introduce an approach in which gravitational field equations are obtained from varying some degrees of freedom other than $g_{ab}$ in a maximization principle. When the new degrees of freedom are varied in the action, 
  the field equations must remain invariant under the shift $\mathcal{L}_{matt}\to \mathcal{L}_{matt}+\lambda_m$ of the matter Lagrangian $\mathcal{L}_{matt}$ by a constant $\lambda_m$. This will give us some kind of `gauge freedom' to absorb any $\lambda_m$ while maintaining general covariance. 

Once we obtain a theory in which gravitational action is invariant under the shift $T_{ab}\to T_{ab}+\Lambda g_{ab}$,  we would have   succeeded in making gravity  decouple from the bulk vacuum energy. \textit{This --- by itself --- is considerable progress}; for example, this will explain why bulk \cc\  --- treated as a parameter in the low energy Lagrangian --- is irrelevant and can be taken to be zero.\footnote{The issue, of course, is related to the fact non-gravitational physics does not care about the absolute zero of energy while gravity does. Curiously, we do not have theoretical formalism of even non-gravitational physics --- say, in standard quantum mechanics  ---  which  is \textit{manifestedly} invariant under shifting of the origin of energy and depends only on energy differences.} Of course, given the observations, there still remains the second issue of explaining the observed value of the \cc. Once the bulk value of the \cc\ (or vacuum energy) decouples from gravity, \textit{classical} gravity becomes immune to \cc; that is, the bulk classical \cc\ can be gauged away.
Any observed value of the \cc\ has to be necessarily a \textit{quantum} phenomenon arising as a relic of microscopic spacetime fluctuations \cite{cc2}. 
 There must exist a deep principle in quantum gravity which leaves its non-perturbative trace even in the low energy limit
that appears as the \cc.

\section{Gravity as an emergent phenomenon}

I will now provide an alternative perspective on gravity \cite{roorkee,aseementropy,grf08} which will lead to field equations which are invariant under the shift $T_{ab}\to T_{ab}+\Lambda g_{ab}$.
I argue  that we have  misunderstood the true nature of gravity because of the way the ideas evolved historically and show that, when seen with the `right side up', the description of gravity becomes remarkably simple and explains features which we never thought needed explanation! 

The historical development of some of the ideas we are interested in is indicated in Table 1.  Einstein started with the Principle of Equivalence and --- with a few thought experiments --- motivated why gravity should be described by a metric  of spacetime. This approach gave the correct backdrop for the equality of inertial and gravitational masses and the \textit{kinematics} of gravity.  But then he needed to write down the field equations which govern the \textit{dynamics} of $g_{ab}$ and here is where the trouble started. 
There is  no good guiding principle which Einstein could use that leads in a natural fashion to $G_{ab}=\kappa T_{ab}$ or to the corresponding action principle (explaining several false starts he had).
Sure, one can obtain them from a series of postulates but they just do not have the same compelling force as, for example, the Principle of Equivalence. 

Nevertheless, we all accepted general relativity, Einstein's equations and their solutions; after all, they  agreed with observations so well!
But --- conceptually --- strange things happen as soon as: (i) we let the metric to be dynamical and (ii) allow for arbitrary coordinate transformations or, equivalently, observers on any timelike curve examining physics.
{\it Horizons are inevitable in such a theory and they are always observer dependent.} This  arises as follows:
(i) Principle of equivalence implies that trajectories of light will be affected by gravity. So in any theory which links gravity to spacetime dynamics, we can have nontrivial null surfaces which block information from certain class of observers. (ii) Similarly, one can construct timelike congruences (e.g., uniformly accelerated trajectories) such that all the curves in such a congruence have a horizon. You can't avoid horizons.
What is more, the horizon is \textit{always} an observer dependent concept, even when it can be given a purely geometrical definition.
For example, the $r=2M$ surface in Schwarzschild geometry  acts operationally as a horizon \textit{only} for the class of observers who choose to stay at $r>2M$ and not for the observers falling into the black hole. 

Once we have horizons --- which  are inevitable --- we get into  more trouble. It is an accepted dictum  that {\it all observers have a right to describe physics using an effective theory \cite{paddy1} based only on the variables (s)he can access.}
(This was, of course, the lesson from renormalization group theory. To describe physics at 10 GeV you shouldn't need to know what happens at $10^{14}$ GeV in "good" theories.)
This raises the famous question first posed by Wheeler to Bekenstein: What happens if you mix cold and hot tea and pour it down a horizon, erasing all traces of ``crime" in increasing the entropy of the world?\footnote{This is based on what Wheeler told me in 1985, from \textit{his} recollection of events; it is also mentioned in his book \cite{wheeler}. I have heard somewhat different versions from other sources.} The answer to such thought experiments  \textit{demands} that
horizons should have an entropy which should increase when energy flows across it.

With hindsight, this is obvious. The Schwarschild horizon --- or for that matter any metric which  behaves locally like Rindler metric --- has a temperature which can be identified by the Euclidean continuation \cite{ghds}. If energy flows across a hot horizon $dE/T=dS$ leads to the entropy of the horizon. Again, historically,  nobody --- including Wheeler and Bekenstein ---  looked  at the   periodicity in the Euclidean time (in Rindler or Schwarzschild metrics) \textit{before} 
Hawking's result came! And the idea of Rindler temperature came \textit{after} that of black hole temperature! 
So in summary, the history proceeded as indicated in Table~\ref{tab:table1}:
\begin{table*}
\hrule
\begin{tabular}{ll}
&\\
&\textbf{Principle of equivalence (Einstein $\sim$ 1908)}\\
$\Rightarrow$&\textbf{Gravity is described by the metric $g_{ab}$ (Einstein $\sim$ 1908)}\\
\textbf{?}&\textbf{Postulate Einstein's equations \textit{without a real guiding principle!} (Einstein $\sim$ 1915)}\\
$\Rightarrow$&\textbf{Black hole solutions with horizons  allowing the entropy of}\\
&\textbf{hot tea to be hidden (Wheeler $\sim$ 1971)}\\
$\Rightarrow$&\textbf{Entropy of black hole horizon (Bekenstein 1972)}\\
$\Rightarrow$&\textbf{Temperature of black hole horizon (Hawking 1975)}\\
$\Rightarrow$&\textbf{Temperature of the Rindler horizon (Davies, Unruh 1975-76)}\\
&\\
\end{tabular}
\hrule
\caption{Conventional perspective of gravity}
\label{tab:table1}
\end{table*}
This historical sequence raises a some serious issues for which there is no satisfactory answer in the conventional approach:

\begin{itemize}
\item 
\textit{How can horizons have temperature without the spacetime having a microstructure?}

They simply cannot. Recall that the 
thermodynamic description of matter at finite temperature provides a crucial window into the existence of the corpuscular substructure of solids. As Boltzmann taught us, heat is a form of motion
and we will not have the thermodynamic layer of description if  matter is a continuum all the way to the finest scale and atoms did not exist! \textit{The mere existence of a thermodynamic layer in the description is proof enough that there are microscopic degrees of freedom.} --- in a solid or in a
 spacetime.   In the conventional approach, we are completely at a loss to understand 
 why horizons are hot  or what kind of `motion' is this `heat'.

To tackle this issue, it is necessary to abandon the usual picture of treating the
metric as  the fundamental dynamical degrees of freedom of the theory and treat it as
 providing a
coarse grained description of the spacetime at macroscopic scales,
somewhat like the density of a solid ---  which has no meaning at atomic  
scales \cite{elastic}.\footnote{The unknown, microscopic degrees of freedom of
spacetime (which should be analogous to the atoms in the case of
solids), should normally play a role only when spacetime is probed at Planck
scales (which would be analogous to the lattice spacing of a solid
\cite{zeropoint}).  So we normally expect the  microscopic structure of spacetime
to manifest itself only at Planck scales or near singularities of the
classical theory. However, in a manner which is not fully understood,
the horizons ---  which block information from certain classes of
observers --- link \cite{magglass} certain aspects of microscopic
physics with the bulk dynamics, just as thermodynamics can provide a
link between statistical mechanics and (zero temperature) dynamics of
a solid. The  reason is probably related to the fact that horizons
lead to infinite redshift, which probes \textit{virtual} high energy
processes; it is, however, difficult to establish this claim in
mathematical terms.} 
 
\item
\textit{Why is it that Einstein's equations reduces to a thermodynamic identity  for virtual displacements of a  horizon?}
 
Here is the first algebraic mystery --- which has no explanation in conventional approach --- suggesting a deep connection between the dynamical equations governing the metric and the thermodynamics of horizons. The first example  was provided in ref.
\cite{paddy2}
 in which it was shown that, in the case of spherically symmetric horizons,  Einstein's equations can be interpreted as a 
thermodynamic relation $TdS=dE+PdV$ arising out of virtual
radial displacements of the horizon. Further work showed that this result is valid in \textit{all} the cases for which explicit computation can be carried out --- as diverse as the
Friedmann models  as well as  rotating and time dependent horizons
in Einstein's theory \cite{rongencai}. Treating them as just some solutions to Einstein's field equations we cannot understand these results.
\item
\textit{Why is Einstein-Hilbert action is holographic with a surface term that encodes same information as the bulk?}

The Einstein-Hilbert Lagrangian has the structure $L_{EH}\propto R\sim (\partial g)^2+ {\partial^2g}$.
In the usual approach the surface term arising from  $L_{sur}\propto \partial^2g$ has to be ignored or canceled to get Einstein's equations from  $L_{bulk}\propto (\partial g)^2$.  
But there is a 
peculiar (again unexplained) relationship between $L_{bulk}$ and $L_{sur}$:
\begin{equation}
    \sqrt{-g}L_{sur}=-\partial_a\left(g_{ij}
\frac{\partial \sqrt{-g}L_{bulk}}{\partial(\partial_ag_{ij})}\right)
\end{equation}
This shows that the gravitational action is `holographic' \cite{cc1,tpholo} with the same information being coded in both the bulk and surface terms making either one of them  to be sufficient.  It is well known that varying $g_{ab}$ in $L_{\rm bulk}$ leads to the standard field  equations. More remarkable is the fact that one can also obtain Einstein's equations from an action principle which  uses only the surface term and the  virtual displacements of horizons \cite{paris}
\textit{without} treating the metric as a dynamical variable. 
\item
\textit{Why does the surface term in Einstein-Hilbert action give the horizon entropy?}

Yet another algebraic result which defies physical understanding! You first throw away the surface term in the action, vary the rest to get the field equations, find a solution with a horizon, compute its entropy --- only to discover that the surface term you threw away is intimately related to the entropy. 
\item
\textit{And, most importantly, why do all these results hold for a much wider class of theories than Einstein gravity,  like \LL models?}

There are more serious  `algebraic accidents' in store.
 Recent work has shown that \textit{all the  thermodynamic features described above extend far beyond Einstein's theory.}
 The connection between field equations and the thermodynamic relation $TdS=dE+PdV$ 
 is not restricted to
Einstein's theory (GR) alone, but is in fact true for the
case of the generalized, higher derivative \LL gravitational theory in
$D$ dimensions as well \cite{aseem-sudipta}. The same is true  for the holographic structure of the action functional \cite{ayan}: the \LL action has the same structure and --- again --- the entropy of the horizons is related to the surface term of the action. 
\end{itemize}

I believe these (and several related features) are not algebraic accidents but indicate that we have been looking at gravity the wrong way around. In the proper perspective, these features should emerge as naturally as the equivalence of inertial and gravitational masses emerges in the geometric description of the kinematics of gravity. 
\textit{These results show that the thermodynamic description is far more general than just Einstein's theory} and occurs in a wide class of theories in which the metric determines the structure of the light cones and null surfaces exist blocking the information.
So instead of the historical path,  I will proceed as in Table 2  reversing  most of the arrows \cite{grf08}:

Let me elaborate. Take an event $P$ and introduce a local inertial frame (LIF) around it with coordinates $X^a$.    Go from the LIF to a local Rindler frame (LRF) coordinates $x^a$ by accelerating along, say, x-axis with an acceleration $\kappa$. This LRF and its local horizon $\mathcal{H}(x=0)$ will exist within a region of size $L\ll\mathcal{R}^{-1/2}$ as long as $\kappa^{-1}\ll\mathcal{R}^{-1/2}$ 
 where $\mathcal{R}$ is a typical component of curvature tensor. 
 Now people can pour tea across $\mathcal{H}$ as suggested by Wheeler. 
Alternatively, one can consider a  virtual displacement of the $\mathcal{H}$ normal to itself
engulfing the tea. Either way, some entropy will be lost to the  outside observers unless the horizon has an entropy. That is, displacing a piece of local Rindler horizon should cost some entropy $S_{grav}$, say. It is then natural to demand that  the dynamics should follow from the prescription
$\delta[S_{grav}+S_{matt}]=0$.

 \begin{table*}
\hrule
\begin{tabular}{ll}
&\\
&\textbf{Principle of equivalence}\\
$\Rightarrow$& \textbf{Gravity is described by the metric $g_{ab}$} \\
$\Rightarrow$& \textbf{Existence of local Rindler frames (LRFs) with  horizons $\mathcal{H}$ around any event} \\
$\Rightarrow$& \textbf{Temperature associated with $\mathcal{H}$ is obtainable from the Euclidean continuation}\\
$\Rightarrow$& \textbf{Virtual displacements of $\mathcal{H}$ allow for flow of energy across a hot horizon hiding}\\
&\textbf{an entropy $dS=dE/T$ as perceived by a given observer}\\
$\Rightarrow$& \textbf{The local horizon must have an entropy, $S_{grav}$} \\
$\Rightarrow$& \textbf{The dynamics should arise from maximizing the total entropy of horizon ($S_{grav}$)}\\
&\textbf{plus matter ($S_m$) for all LRF's \textit{without} varying the metric}\\
$\Rightarrow$& \textbf{The field equations are those of \LL gravity with Einstein's gravity} \\
& \textbf{emerging as the lowest order term} \\
$\Rightarrow$& \textbf{The theory is invariant under the shift $T_{ab}\to T_{ab}+\Lambda g_{ab}$
allowing}\\
& \textbf{the bulk \cc to be `gauged away'}\\ 
&\\
\end{tabular}
\hrule
\caption{Alternative perspective of gravity}
\end{table*}

All we need is the expressions for $S_{matt}$ and $S_{grav}$. 
I will now write down the general expressions for both \cite{aseementropy,grf08}, such that they have the correct interpretation in the LRF. Assuming a $D(\ge 4)$ dimensional spacetime for the 
later convenience, I take:
\begin{equation}
S_{grav}= - 4\int_\Cal{V}{d^Dx\sqrt{-g}}
    P_{ab}^{\ph{a}\ph{b}cd} \D_cn^a\D_dn^b;\qquad
S_{\rm matt}=\int_\Cal{V}{d^Dx\sqrt{-g}}
      T_{ab}n^an^b
\label{Sgravmatt}
\end{equation} 
where $n^a$ is vector field which will reduce to the null normal [$\partial_a (T-X)]$ on the horizon $\mathcal{H}$, $T_{ab}$ is the matter energy momentum tensor and $P_{ab}^{\ph{a}\ph{b}cd}$ is defined below. 
For mathematical convenience, it is better to treat $n_a$ as a vector field with a fixed norm $n_an^a\equiv\epsilon$  rather than as a strictly null vector. This allows us to obtain the horizon as a limiting case of a `stretched horizon' which is a
 timelike surface. [This, as well as some other subtleties in the variational principle are described in ref. \cite{aseementropy}.]
The  $S_{\rm matt}$ is easy to understand. In the LRF, (with $-g_{tt} = 2\kappa x =  g^{xx}, \sqrt{-g}=1$) an infinitesimal spacetime region  will contribute $T_{ab} n^bn^bd^3x\ dt=\delta E \ dt$ which on integration over $t$ in the range $(0,\beta)$
where $\beta^{-1}=T = (\kappa/2\pi)$  gives
$
\delta S_{matter}=\beta\delta E=\beta T_{ab} n^a n^b d^3x
$ 
when the energy flows across a surface with normal $n^a$. 
 Integrating,  we get $S_{\rm matt}$ to which the expression in Eq.~(\ref{Sgravmatt}) reduces to in LRF.
 (For example, if $T_{ab}$ is due to an ideal fluid at rest in LIF, $T_{ab}n^an^b$ will contribute $(\rho+P)$, which --- by Gibbs-Duhem relation --- is just $Ts$ where $s$ is the entropy density. Integrating over $\sqrt{-g} d^4 x = dt d^3 x$ with $0<t<\beta$ gives $S_{\rm matt}$.)

The $S_{grav}$, on the other hand, is a general quadratic functional of the derivatives of $n_a$ which is the form of entropy of an elastic solid, say, if $n^a$ is the displacement field.
Here we interpret it as the entropy cost for virtual displacement of horizon. The crucial requirement is that, dynamics for the \textit{background spacetime} should emerge when we set $(\delta S_{tot}/\delta n_a)$=0 for \textit{all} null vectors $n^a$
(rather than an equation for $n^a$). Incredibly enough, this can be achieved if (and only if) (i) the tensor $P_{abcd}$ 
has the algebraic symmetries similar to the Riemann tensor $R_{abcd}$
 and (ii) we have
$
\D_{a}P^{abcd}=0=\D_{a}T^{ab}$.
%\label{ent-func-1}
One can now show that  \cite{aseementropy} such  a tensor can be constructed as a 
a series in the powers of the derivatives of the metric:
\begin{equation}
P^{abcd} (g_{ij},R_{ijkl}) = c_1\,{P}_{(1)}{}^{abcd} (g_{ij}) +
c_2\, {P}_{(2)}{}^{abcd} (g_{ij},R_{ijkl})  
+ \cdots \,,
\label{derexp}
\end{equation} 
where $c_1, c_2, \cdots$ are coupling constants with the \textit{unique} m-th order term
being
$
{P}{}_{ab}^{\ph{a}\ph{b}cd}\propto
\partial\LDm/\partial R^{ab}_{{\ph{ab}cd}} 
$
where $\LDm$ is the m-th order \LL Lagrangian \cite{paris,lovelock}. 
Then maximizing ($S_{\rm grav} + S_m$) gives \cite{aseementropy}:
\begin{equation}
16\pi\left[ P_{b}^{\ph{b}ijk}R^{a}_{\ph{a}ijk}-\frac{1}{2}\delta^a_b\LDm\right]=
 8\pi T{}_b^a +\Lambda\delta^a_b   
\label{ent-func-71}
\end{equation}
These are identical to the field equations for \LL gravity with a cosmological constant arising as an undetermined integration constant. 
The lowest order term 
$
{P}{}^{ab}_{cd} =(1/32\pi)
(\delta^a_c \delta^b_d-\delta^a_d \delta^b_c)
$ leads to Einstein's theory
while the first order term  gives the  Gauss-Bonnet correction.
One can show, in the general case of \LL\ theory, Eq.~(\ref{Sgravmatt}) \textit{does} give the correct gravitational entropy justifying our choice.
Remarkably enough, we can derive not only Einstein's theory but even \LL theory from a dual description in terms on the normalized vectors in spacetime, \textit{without varying $g_{ab}$ in an action functional!}
\footnote{Incidentally, this solves another problem related to \LL gravity. Since the standard action funtional for \LL gravity contains second derivatives of the metric in a non-trivial manner --- expect for Einstein's gravity in which the action is linear in second derivatives --- the variational principle is \textit{not} well-defined. It is not known, except in the case of Einstein's theory and in case of GB gravity, what counterterms are to be added to the action to make the variational principle well-defined. We completely bypass this problem in our approach since the metric is not varied.}

The crucial feature of the coupling between matter and gravity through $T_{ab}n^an^b$ 
is that, under the shift $T_{ab}\to T_{ab}+\rho_0g_{ab}$, the $\rho_0$  term in the action in \eqn{Sgravmatt} decouples from $n^a$ and becomes irrelevant:
\begin{equation}
\int_\Cal{V}{d^Dx\sqrt{-g}}T_{ab}n^an^b \to 
\int_\Cal{V}{d^Dx\sqrt{-g}} T_{ab}n^an^b +
\int_\Cal{V}{d^Dx\sqrt{-g}}\epsilon\rho_0
\end{equation} 
Since $\epsilon=n_an^a$ is not varied when $n_a$ is varied there is no coupling between $\rho_0$ and the dynamical variables $n_a$ and the theory is invariant under the shift  $T_{ab}\to T_{ab}+\rho_0g_{ab}$. Of course, when $n_a$ is null --- in the limit of stretched horizon becoming the horizon --- the second term vanishes.
 We see that just the condition $n_an^a=$ constant on the  dynamical variables have led to a `gauge freedom' which allows an arbitrary integration constant to appear in the theory which can absorb the bulk cosmological constant.

\section{Surface degrees of freedom and the observed value of \cc}

The description of gravity  given above provides a natural backdrop for gauging away the bulk value of the cosmological constant since it decouples from the dynamical degrees of freedom in the theory.  Once the bulk term is eliminated, 
what is observable through gravitational effects, in the correct theory of quantum gravity, should be the \textit{fluctuations} in the vacuum energy.
These fluctuations will be non-zero if the universe has a DeSitter horizon which provides a confining 
volume. In this paradigm the
 vacuum structure can readjust  to gauge away the bulk energy density $\rho_{_{\rm UV}}\simeq L_P^{-4}$ while quantum \textit{fluctuations} can generate
the observed value $\rho_{\rm DE}$. This  boils down to explaining the numerical value of 
 $L_\Lambda/L_P(\simeq\exp(\sqrt{2}\pi^4))\approx 10^{60}$. \textit{I do not yet know how to do this} but will provide arguments as to why the approach with surface degrees of freedom might hold the key.

I shall  argue that, the quantum fluctuations that lead to the observed value of \cc\ should scale with the surface area rather than the bulk volume to give the correct answer. This, in turn, suggests that
the
 relevant degrees of freedom (which replaces the metric) should be linked to  surfaces in spacetime rather than
bulk regions --- something we have already seen in the emergent gravity approach.
The observed \cc\ should then be a  relic of quantum gravitational physics and should arise from  degrees of freedom which scale as the surface area. 

 Consider a 3-dimensional region of size $L$ with a bounding area which scales as $L^2$. Let us assume that we
 associate   with
 this region  $N$ microscopic cells of size $L_P$ 
each having a Poissonian fluctuation in energy of amount $E_P\approx 1/L_P$. Then the mean square fluctuation of energy in this region will be $(\Delta E)^2\approx NL_P^{-2}$ corresponding to the energy density
$\rho=\Delta E/L^3=\sqrt{N}/L_PL^3$. If we make the usual assumption that $N=N_{vol}\approx (L/L_P)^3$, this will give
\begin{equation}
\rho=\frac{\sqrt{N_{vol}}}{L_PL^3}=\frac{1}{L_P^4}\frab{L_P}{L}^{3/2} \quad \text {(bulk\ fluctuations)}
\end{equation} 
On the other hand, if we assume that  the relevant degrees of freedom scale as the surface area of the region, then $N=N_{sur}\approx (L/L_P)^2$
and the relevant energy density is
\begin{equation}
\rho=\frac{\sqrt{N_{sur}}}{L_PL^3}=\frac{1}{L_P^4}\frab{L_P}{L}^2=\frac{1}{L_P^2L^2} \quad \text {(surface\ fluctuations)}
\label{sur}
\end{equation}
If we take $L\approx L_\Lambda$, the surface fluctuations  give precisely the geometric mean, of the two energy scales
 $\rho_{_{\rm UV}}=1/L_P^4$ and $\rho_{_{\rm IR}}=1/L_\Lambda^4$ in natural units ($c=\hbar=1$), which is the
observed value of the energy density contributed by the \cc. On the other hand, the bulk \textit{fluctuations} lead to an energy density which is larger by a factor 
$(L/L_P)^{1/2}$. 
 Of course, if we do not take fluctuations in energy but coherently add them, we will get $N/L_PL^3$ which is $1/L_P^4$ for the bulk and $(1/L_P)^4(L_P/L)$
for the surface. In summary, we have the following hierarchy:
\begin{equation}
\rho=\frac{1}{L_P^4}\times \left[1,\frab{L_P}{L},
\frab{L_P}{L}^{3/2},
\frab{L_P}{L}^2,
\frab{L_P}{L}^4 .....\right]
\end{equation} 
in which the first one arises by coherently adding energies $(1/L_P)$ per cell with
$N_{vol}=(L/L_P)^3$ cells; the second arises from coherently adding energies $(1/L_P)$ per cell with
$N_{sur}=(L/L_P)^2$ cells; the third one is obtained by taking \textit{fluctuations} in energy and using $N_{vol}$ cells; the fourth from energy fluctuations with $N_{sur}$ cells; and finally the last one is the thermal energy of the DeSitter space if we take $L\approx L_\Lambda$;  clearly the further terms are irrelevant due to this vacuum noise. 
Of all these, the only viable possibility is what arises if we assume that: 
(a)
The number of active degrees of freedom in a region of size $L$ scales as $N_{sur}=(L/L_P)^2$.
(b)
It is the \textit{fluctuations} in the energy that contributes to the cosmological constant \cite{cc1,cc2} and the bulk energy does not gravitate.

The role of energy fluctuations contributing to gravity also arises, more formally, when we study the question of \emph{detecting} the energy
density using gravitational field as a probe.
 Recall that a detector with a linear  coupling  to the {\it field} $\phi$
actually responds to $\langle 0|\phi(x)\phi(y)|0\rangle$ rather than to the field itself \cite{probe}. Similarly, one can use the gravitational field as a natural ``detector" of energy momentum tensor $T_{ab}$ with the standard coupling $L=\kappa h_{ab}T^{ab}$. Such a model was analyzed in detail in ref.~\cite{tptptmunu} and it was shown that the gravitational field responds to the two point function $\langle 0|T_{ab}(x)T_{cd}(y)|0\rangle $. In fact, it is essentially this fluctuations in the energy density which is computed in the inflationary models \cite{inflation} as the  {\it source} for gravitational field, as stressed in
ref.~\cite{tplp}. All these suggest treating the energy fluctuations as the physical quantity ``detected" by gravity, when
one  incorporates quantum effects.  

Quantum theory, especially the paradigm of renormalization group has taught us that the  concept of the vacuum
state  depends on the scale at which it is probed. The vacuum state which we use to study the
lattice vibrations in a solid, say, is not the same as vacuum state of the QED
 and it is not appropriate to ask questions about the vacuum without specifying the scale. 
If the \cc\ arises due to the fluctuations in the energy density of the vacuum, then one needs to understand the structure of the quantum gravitational vacuum at cosmological scales. 
 If the spacetime has a cosmological horizon which blocks information, the natural scale is provided by the size of the horizon,  $L_\Lambda$, and we should use observables defined within the accessible region. 
The operator $H(<L_\Lambda)$, corresponding to the total energy  inside
a region bounded by a cosmological horizon, will exhibit fluctuations  $\Delta E$ since vacuum state is not an eigenstate of 
{\it this} operator. A rigorous calculation (see the first reference in \cite{cc2}) shows that 
  the fluctuations in the energy density of the vacuum in a sphere of radius $L_\Lambda$ 
 is given by 
 \begin{equation}
 \Delta \rho_{\rm vac}  = \frac{\Delta E}{L_\Lambda^3} \propto L_P^{-2}L_\Lambda^{-2} 
 \label{final}
 \end{equation}
 The numerical coefficient will depend on $c_1$ as well as the precise nature of infrared cutoff 
 radius;
 but it is a fact of life that a fluctuation of magnitude $\Delta\rho_{vac}\simeq H_\Lambda^2/G$ will exist in the
energy density inside a sphere of radius $H_\Lambda^{-1}$ if Planck length is the UV cut off. 
On the other hand, since observations suggest that there is a $\rho_{vac}$ of similar magnitude in the universe it seems 
natural to identify the two. Our approach explains why there is a \textit{surviving} cosmological constant which satisfies 
$\rho_{_{\rm DE}}=\sqrt{\rho_{_{\rm IR}}\rho_{_{\rm UV}}}$.
  
It is, of course, possible to give all kinds of arguments (mostly based on some version of `holography') to motivate an expression like \eq{final}. In most of these approaches, $L_\Lambda$ will be identified with a \textit{time dependent} length scale in the Friedmann universe --- Hubble radius, past horizon, future horizon .... --- and one will try to see whether the resulting model agrees with observations. There are two issues I want to briefly discuss in this context. 

First, I am not sure this will lead to a satisfactory solution; it could very well be that some unknown  quantum gravitational effect will actually give the ratio $L_\Lambda/L_P\approx 10^{61}$ just as we expect some theory to eventually tell us why $m_\nu/M_{Pl}\approx 10^{-30}$. (I have already indicated a non-perturbative numerology: $L_\Lambda/L_P=\exp(\sqrt{2}\pi^4)\approx 10^{60}$!)
Cosmological constant    is then interpreted as small because it is a nonperturbative quantum relic.

Second, and more important, I stress that invoking \eq{final} with some cosmological length scale for $L_\Lambda$  is completely meaningless in the  models of gravity in which the metric couples to the bulk energy density. Until we have a paradigm in place which allows us to ignore the bulk \cc\ --- which most of the ad hoc, holographic dark energy type models do not have --- one cannot invoke such a procedure. This is particularly true in any model which leads to \eq{final} through any kind of quantum fluctuation. All such approaches will require  a UV cut-off (at Planck scale) to give a finite answer; once it is imposed, one will always get a bulk contribution $\rho_{UV}\approx L_P^{-4}$ with  the usual problems. It is only because \textit{we} have a way of decoupling the bulk term  from contributing to the dynamical equations that, we have a right to look at the subdominant term $L_P^{-4}(L_P/L_\Lambda)^2$. Approaches in which the sub-dominant term is introduced by an ad hoc manner are conceptually flawed since the bulk term cannot be ignored in these usual approaches to gravity.
Getting the correct value of the cosmological constant from the energy fluctuations is not as difficult as understanding why the bulk value  (which is larger
by $10^{120}$!) can be ignored. Our approach provides a natural framework for 
ignoring the bulk term --- and as a bonus --- the \textit{possibility} of obtaining the right value for the cosmological 
constant from the fluctuations.

\section{Conclusions}

 The simplest choice for the negative pressure component in the universe is the cosmological constant; other models based on scalar fields (as well as those based on branes etc. which I  have not discussed) do not alleviate the difficulties faced by \cc\  and --- in fact --- makes them worse. I have shown that it is impossible to solve the \cc\ problem unless the gravitational sector of the theory is invariant under the shift $T_{ab}\to T_{ab}+\lambda_mg_{ab}$. Any approach which does not address this issue cannot provide a comprehensive solution to the \cc\ problem.
 But general covariance requires us to use the measure $\sqrt{-g}d^Dx$ in D-dimensions
in the action which will couple the metric (through its determinant) to the matter sector. Hence, as long as we insist on metric as the fundamental variable that is varied in an action principle, one cannot address
this issue.  So we need to introduce some other degrees of freedom and an effective action which, however, is capable of constraining the background metric.

An action principle, based on the  normalized vector fields in spacetime, satisfies all these criteria mentioned above. The new action does not couple to the bulk energy density and maintains invariance under the shift $T_{ab}\to T_{ab}+\lambda_mg_{ab}$. What is more, the on-shell value of the action is related to the entropy of horizons showing the relevant degrees of freedom scales as the area of the bounding surface.
 
 Since our formalism ensures that the bulk energy density does not contribute to gravity --- \textit{and only because of that} --- it makes sense to compute the next order correction due to fluctuations in the energy density. I have not been able to  compute this rigorously with the machinery available, but a plausible case can be made as how this approach might  lead to the correct, observed, value of the \cc.

 \end{document}